# First Principle Study of Magnetism and Magneto-structural Coupling in Gallium Ferrite

Amritendu Roy[1], Rajendra Prasad[2], Sushil Auluck[3], and Ashish Garg[1*]

[1] Department of Materials Science & Engineering, Indian Institute of Technology, Kanpur - 208016, India
[2] Department of Physics, Indian Institute of Technology, Kanpur - 208016, India
[3] National Physical Laboratory, Dr K S Krishnan Marg, New Delhi-110012, India

**ABSTRACT**

We report a first-principles study of the magnetic properties, site disorder and magneto-structural coupling in multiferroic gallium ferrite (GFO) using local spin density approximation (LSDA+U) of density functional theory. The calculations of the ground state A-type antiferromagnetic structure predict magnetic moments consistent with the experiments whilst consideration of spin-orbit coupling yields a net orbital moment of ~ 0.025 $\mu_B$/Fe site also in good accordance with the experiments. We find that though site disorder is not spontaneous in the ground state, interchange between Fe2 and Ga2 sites is most favored in the disordered state. The results show that ferrimagnetism in GFO is due to Ga-Fe site disordering such that Fe spins at Ga1 and Ga2 sites are antiferromagnetically aligned while maintaining ferromagnetic coupling between Fe spins at Ga1 and Fe1 sites as well as between Fe spins at Ga2 and Fe2 sites. The effect of spin configuration on the structural distortion clearly indicates presence of magneto-structural coupling in GFO.

---

[*] Corresponding author, Tel: +91-512-2597904; FAX - +91-512-2597505, E-mail: ashishg@iitk.ac.in

# I. INTRODUCTION

In the recent years, magnetoelectric (ME) and multiferroic (MF) materials have generated immense research interest owing to their potential in novel device applications such as in sensors, transducers and data storage. [1-3] While experiments have been performed on a few selected materials, first principle studies have proven useful in prediction of new materials[4-6] and coupling mechanisms [7, 8]. As various theoretical and experimental studies have shown that magnetism in the ME materials, such as antiferromagentic $BiFeO_3$ [9] can be modified by doping [10-12], reducing particle size [13] or by application of epitaxial strain [14, 15]. Moreover, the ME coupling is shown to be greatly affected by epitaxial strain, pointing towards large magnetostructural (MS) coupling as it has been predicted that the materials with large MS coupling can even undergo antiferro-ferromagnetic transition upon application of suitable amount of epitaxial strain.[16] MS coupling in another important class of materials, antiferromagnetic rare earth manganites with giant ME coupling, over and above ME coupling, induces additional change in the electric dipole moment resulting in large spontaneous polarization.[17]

In this context, gallium ferrite ($GaFeO_3$ or GFO) emerges as an interesting material simultaneously exhibiting piezoelectricity and ferrimagnetism at temperatures near room temperature (RT) as its magnetic transition temperature can be tuned close to or above RT depending upon Ga:Fe ratio [18, 19] and processing condition.[18-21] GFO has an orthorhombic structure with *$Pc2_1n$* symmetry which is retained over a wide temperature range of 4-700 K. [18, 22] First-principles studies have shown that GFO is an antiferromagnet at 0 K, [23, 24] but cation site disorder driven by almost similar ionic sizes of Ga and Fe results in uneven distribution of magnetic moments in the octahedral interstitial sites and is believed to render GFO as a ferrimagnetic material with substantial magnetic moment below $T_c$ [18]. GFO, reported to possess strong ME coupling [18, 25], is also likely to exhibit a cross-linking between the structural and magnetic order parameters. Therefore, a study of magnetic behavior focusing on the effect of structural distortion on its magnetic behavior would be imperative to explore this further. Moreover, except a few reports[18, 24] with rather different emphasis, the studies on the fundamental understanding of microscopic magnetic behavior of GFO such as effect of site disorder on the magnetic moments and magnetic structure are severely lacking.

In this paper, we present a detailed study on the atomic scale magnetic behavior of GFO including the effect of site disorder and possible existence of MS coupling in the ground state. The calculations indicate presence of MS coupling in GFO. We also find that the magnetism is solely due to cation site disorder in GFO. The calculated magnetic moments of Fe ions in GFO agree well with, albeit scarce, previously reported experimental [18] and theoretical [24] data. Although, there are many reports of magneto-structural coupling in materials, the phenomenon is not well understood at a microscopic level. A first principles calculation affords us this opportunity to explore the origin of MS coupling in GFO.

## II. CALCULATION DETAILS

We employed first principles density functional theory [26] based Vienna ab-initio simulation package (VASP) [27] with projector augmented wave method (PAW) [28] in our calculation. Local spin density approximation (LSDA+U) [29] with Hubbard parameter, U = 5 eV, and the exchange interaction, J = 1 eV was used to solve the Kohn-Sham equation. [30] The value of U was so determined that the calculated magnetic moments of Fe ions agree well with the experimentally determined moments. Small variation of U was found not to affect the system's stability. Calculations are based on the stoichiometric GFO with no partial occupancies of the cations. We included 3 valence electrons of Ga ($4s^2 4p^1$), 8 for Fe ($3d^7 4s^1$) and 6 for O ($2s^2 2p^4$) ions. The effect of 3d semicore states of Ga ion was found not to affect the structural and magnetic characteristics significantly and therefore, ignored. [23] A plane wave energy cut-off of 550 eV was used. Conjugate gradient algorithm [31] was used for the structural optimization. Structural calculations were performed at 0 K with Monkhorst-Pack [32] 7×7×12 mesh while site-disorder and MS coupling study was performed with 4×4×4 mesh which hardly made differences with that of 7×7×12 mesh calculations. We also repeated some of our calculations using generalized gradient approximation (GGA+U) with the optimized version of Perdew-Burke-Ernzerhof functional for solids (PBEsol) [33] to check the robustness of our LSDA+U calculations. Magnetic measurement of the experimentally synthesized sample [23] was done using Lakeshore vibration sample magnetometer.

## III. RESULTS AND DISCUSSION

### A. Crystal and Magnetic Structures

Ground state crystal structure, determined in our earlier study[23], predicted orthorhombic *Pc2₁n* symmetry with A-type antiferromagnetic spin configuration. Calculated ground state lattice parameters, using LSDA+U, are: *a* = 8.6717 Å, *b* = 9.3027 Å and *c* = 5.0403 Å which correspond well with experiments. Calculated ionic positions, cation-oxygen and cation-cations bond lengths are also in good agreement with the experiments.[23]

Based on the ground state structural data, [23] (see Fig. 1) we calculated Fe-O-Fe bond angles which can be correlated with the super-exchange interaction between O and neighboring $Fe^{3+}$ ions. Generally, larger the Fe-O-Fe bond angle, stronger is the antiferromagnetic super-exchange. [34, 35] From the structural parameters, obtained from our first principles calculations and from the experimental XRD data,[23] we calculated cation-oxygen-cation bond angles as compiled in Table 1. The maximum value of Fe1-O1-Fe2, bond angle is ~168.54° while other angles are: Fe1-O3-Fe2, 123.13° and Fe1-O5-Fe2, 126.23°, respectively calculated using LSDA+U method as observed in Table1. As a check, similar values were also obtained using GGA+U also. Situated at the next nearest neighbor positions, O4 and O6 also form wide Fe-O-Fe bonds with angles Fe1-O4-Fe2, 143.36° and Fe1-O6-Fe2, 174.46°. As we show in the following paragraph, such large Fe-O-Fe bond angles (larger than 90°) lead to noticeable super-exchange interaction between Fe and O ions which is reflected in significantly large magnetic moments of O. Maximum bond angle among Fe-O-Ga is observed for Fe1-O1-Ga2 ~166.08°. Any Fe ion that occupies Ga2 site due to site disorder would therefore form a strong antiferromagnetic spin arrangement with Fe at Fe1 site through super-exchange interaction. This strongly indicates that Fe at Ga2 site would be ferromagnetically aligned with Fe2 ions. This emphasizes that Fe2 ion is also antiferromagnetically coupled to Fe at Ga1 site since the Ga1-O6-Fe2 bond angle, ~124.12° is significantly larger than 90°.[41, 42] It is therefore, reasonable to state that any Fe ion occupying Ga1 site due to site disorder would align itself antiferromagnetically with Fe2 and Fe at Ga2 site and would be antiferromagnetically coupled with Fe1 site.

As discussed in detail in our previous work (Ref. 23), we started our calculation to determine the ground state structure of GFO with four possible antiferromagnetic spin structures, namely, AFM1 (A-type antiferromagnetic), AFM-2 (C-type antiferromagnetic), AFM-3 (G-type antiferromagnetic) and AFM-4 and established that the ground state magnetic structure of GFO is A-type antiferromagnetic (A-AFM) [23]. Now, we calculate the magnetic moments of the constituent ions in the ground state and compare these with the data reported in the literature (see Table 2). Our calculations using LSDA+U method show that Fe1 and Fe2 ions have magnetic moments of + 4.05 $\mu_B$ and - 4.04 $\mu_B$, respectively whereas GGA+U calculations predict Fe magnetic moments of + 4.12 $\mu_B$ and -4.12 $\mu_B$ for Fe1 and Fe2, respectively. We find that while the magnitude of moments agrees well with the experimental data, the sign, though in agreement with the theoretical data shown by Han *et al.* [24], is opposite to the experimental neutron diffraction results showing -3.9 $\mu_B$ and + 4.5 $\mu_B$ for Fe1 and Fe2 respectively. The difference in the sign of the magnetic moments with respect to those obtained from neutron studies is due to an equivalent spin structure which is mirror image of AFM-1. We calculated the total energy of such a structure and found that the energy per unit-cell is identical for either of the two structures. As discussed earlier, magnetic moments manifested by oxygen ions are due to super-exchange interactions with the surrounding Fe ions. Table 2 also shows that moments calculated by Han *et al.* [24] using LSDA+U (without SOC) yielded somewhat larger values of magnetic moments than either the experimental data or the values shown by our results.

Further, to probe the spin-orbit interaction in the ground state structure, we performed spin-orbit coupling in conjunction with LSDA+U calculation assuming spin moment directions of Fe1 and Fe2 to be [001], as suggested by the experiments [36] and the calculations yielded the orbital magnetic moment of ~ 0.025 $\mu_B$/Fe site. This value is slightly larger than 0.02 $\mu_B$/Fe site calculated by Han *et al* [24] and 0.017 $\mu_B$/Fe site measured experimentally at 190 K, by Kim *et al* [36]. Interestingly, while SOC calculations by Han *et al* [24] reduced the magnetic moment by ~0.75 $\mu_B$, our SOC calculation did not alter the magnetic moments significantly. Since our LSDA+U calculations without the consideration of spin-orbit interaction satisfactorily describe the magnetic structure with respect to the experimental results and a more precise magnetic structure would probably not improve the accuracy of our further calculations in a significant way, we ignored spin-orbit interaction in further calculations.

## B. Cation Site Disorder and it's Effect on Magnetic Behaviour

Previous experiments suggest that the magnetic structure of GFO can be influenced significantly by cation site-disorder *i.e.* mixed occupancies of Ga and Fe on each other's sites. Experiments using both neutron [18] and x-ray diffraction [22] techniques reveal that structure of GFO exhibits cation site disorder *i.e.* some of the Ga sites are occupied by Fe ions. Most previous studies indicate that Fe occupation of Ga1 sites is much smaller than Ga2 sites.[18] Experimental study[18] and our previous discussion in section III (A) show, ferromagnetic coupling of Fe at Fe2 and Ga2 sites [18] and thus it is believed that Fe2 ions mostly occupy Ga2 sites. However, there is no concrete evidence supporting this since Fe ions upon getting out of its original sites may change its spin configuration. Since the results shown in the previous paragraphs are based upon the assumption of full site occupancy, we further investigated the cation site disorder in GFO, to determine which Fe ions preferentially occupy Ga sites.

To study the effect of cation site disorder on the magnetic structure, we selectively interchanged Fe and Ga sites and computed total energy of the system. Since, GFO unit-cell contains four ions of each type of cation, such an interchange would result in ¼$^{th}$ site occupancy of Fe ions at Ga sites and *vice-versa*. The change in the energy of the unit-cell with respect to the ground state upon site interchange is plotted in Fig. 2. The figure shows that at 0 K, partial site occupancy is not favored in the ground state, also observed previously by Han *et al.*[24] Therefore, thermal energy and lattice defects are the only likely sources to induce the experimentally observed site disorder in GFO. However, Fig. 2 also shows that among various possible cases of site disorders, Fe2 ions preferentially occupy Ga2 sites is most probable since $\Delta E$, the energy difference with respect to the ground state in that case is minimum. Although these energy differences may be affected by the computational methodology, interestingly, the magnitude of the available thermal energy at room temperature (kT ~25 meV) is of the order of the energy difference for Fe2-Ga2 site disorder indicating towards the role of thermally originated defects.

An important implication of the inclusion of cation site disorder in the calculation would be on the modification of the local magnetic moments. It was observed that upon interchanging Fe1 and Ga1 sites, the average magnetic moment of Fe ion at Ga1 site becomes 3.99 $\mu_B$. While the average magnetic moments of Fe1 ions remain the same, the moments at Fe2 sites are slightly modified with respect to the perfect structure. On the other hand, the magnetic moment of Fe ion at Ga2 site becomes 4.11 $\mu_B$ when Fe2 and Ga2 sites are interchanged. Though, earlier Neutron study [18] shows an increase of ~4.5% in the magnetic moment of Fe2 ion located at Ga2 (4.7 $\mu_B$) site over the original Fe2 magnetic moment (4.5$\mu_B$), our calculations show only ~ 1.5 % increase of the moment of Fe ions at Ga2 site with respect to the parent Fe2 site. Such a small increase in the magnetic moment is probably attributed to the limitation of local density approximation. With Fe2-Ga2 site interchange, the magnetic moments at Fe1 and Fe2 sites are also modified slightly with average moments being + 4.05 $\mu_B$ and - 4.06 $\mu_B$. A closer investigation shows that the Fe1 sites nearest to the occupied Ga2 site have a moment of ~ 4.11 $\mu_B$ while Fe1 site farthest to the Ga2 site has a moment of ~ 4.03 $\mu_B$.

However, the magnetic moments of the Fe2 sites are almost similar to each other. Further, site interchange between Fe1-Ga2 and Fe2-Ga1 also demonstrate changes in local magnetic moments similar to the above two most probable situations of site disorder *i.e.*, Fe2-Ga2 and Fe1-Ga1

As we have shown earlier, large bond angles for Fe-O-Fe demonstrate antiferromagnetic coupling of the spins at Fe1 and Fe2 sites. Further, any Fe at Ga1 site would have ferromagnetic interaction with Fe at Fe1 site while Fe at Ga2 site would make ferromagnetic coupling with Fe at Fe2 site. However, while the ground state structure showed perfect antiferromagnetism with net magnetic moment ~ 0 $\mu_B$, our VSM measurements on crushed single crystals of GFO showed a net magnetic moment of 0.21 $\mu_B$/ Fe site at 120 K. Using the partial site occupancies from the Rietveld refinement data[23] and taking the magnetic moments from the preceding paragraph for different cation sites, we estimated net magnetic moment of 0.24 $\mu_B$/ Fe site which agrees quite well with our experimental results. Therefore, we conclude that ferrimagnetism in GFO is solely due to site disorder in the structure. As noted in earlier paragraphs, the Fe ions at Ga sites would develop different magnetic moments due to different crystal environments than their parent sites.

## C. Determination of Magneto-structural (MS) Coupling

Since GFO is a piezoelectric and antiferromagnetic in the ground state, it is likely to demonstrate piezomagnetism and would possibly have significant MS coupling. To the best of our knowledge, there is no theoretical or experiments work in the literature on these features of GFO. As can be expected, MS coupling in several magnetic compounds results in structural distortion.[37] Depending upon the strength of the coupling, this distortion can even cause structural phase transition as observed in MnO,[38] CrN [37, 39] and LaMnO$_3$ [40], SrMnO$_3$.[16] In this section, to investigate the presence of MS coupling in GFO, we first look at the effect of spin ordering which results in significant stress in the structure and then we demonstrate that structural distortion results in variation in magnetization of the magnetic ions of the system.

We keep AFM-1 as the reference magnetic structure and change the spin configurations according to AFM-2, AFM-3 and AFM-4 structures while maintaining the lattice parameters and ionic positions of the optimized AFM-1 structure. On this basis, we calculated the forces on the atoms and stresses along the *b*-axis and the results are listed in Table 3. Here we observe that the stress, maximum force and hydrostatic pressure on AFM-2, AFM-3 and AFM-4 structures are different indicating that the spin configuration can influence the structural stability and in turn suggests towards the existence of MS coupling in GFO. Subsequently, we relaxed the structures corresponding to the above spin configurations such that the forces on atoms and hydrostatic pressure are close to zero and plotted the lattice parameters and magnetic moment of Fe ion for all the spin configurations, mentioned above, in the respective optimized structures as shown in in Fig. 3 (a) and (b). Here, we see that that not only significant lattice distortion is invoked upon variation in spin configuration; also the magnetic moments at the two Fe sites are affected by the change in the spin configuration. This, therefore, again indicates towards the presence of MS coupling in GFO.

Wojdeł and Íñiguez [15] showed that a quantitative measure of MS coupling can be provided by frozen ion piezomagnetic stress tensor ($\bar{h}_{ij} = \left.\frac{\delta M_i}{\delta \eta_j}\right|_u$) and magnetization change driven by atomic displacement ($\zeta_{pi} = -\Omega \left.\frac{\delta M_i}{\delta u_p}\right|_\eta$), defined at zero external electric field ($M$: magnetization, $\eta$: strain, $u$: displacement). However, in a more simplistic manner, one can argue, if the magnetization of a system is affected by application of an external stress (or vice-versa), the coupling coefficient ($\tau = \frac{dM}{d\sigma}$, σ: applied hydrostatic stress resulting in volumetric strain in the system) would be a measure of MS coupling. Here, we allow the ions within the unit-cell to completely relax in order to accommodate the external stress. In this process, one can combine the effects of structural strain and atomic displacement [15] on the resultant magnetization.

On this basis, we sequentially applied hydrostatic stress (pressure) on the structure and calculated resultant magnetic moments of Fe1 and Fe2 ions and results are shown in Fig. 4. Total energy calculation of the two AFM structures, AFM-1 (A-AFM) and AFM-2 (C-AFM) (stability wise C-AFM is second most stable structure after ground state A-AFM structure[23]) shows no change in the magnetic structure upon application of external pressure (see inset of Fig. 4). However, magnetic moments at Fe1 and Fe2 sites as shown in Fig. 4 demonstrate a small change. Linear fitting of magnetization plotted as a function of external pressure, yields the coefficient of MS coupling (τ) of the order of ~ $2\times10^{-3}$ μ$_B$/GPa. To substantiate the above calculation of τ, we applied the same approach to cubic SrMnO$_3$, a well proven material demonstrating epitaxial strain driven structural (as well as magnetic) phase transition [16] indicating the presence of strong MS coupling. We find that similar calculations on SrMnO$_3$ demonstrate τ ~ $6.5\times10^{-3}$ μ$_B$/GPa and τ ~ $2.18\times10^{-2}$ μ$_B$/GPa in the G-type antiferromagnetic and ferromagnetic states of SrMnO$_3$ respectively. These values being noticeably larger than that seen in GFO, we can deduce that the MS coupling in GFO is relatively weaker and therefore may not drive the system to any structural transition. This is further supported by the absence of any significant structural anomaly near the magnetic transition temperature of GFO.[18, 22] While there are no experimental results to compare the values of MS coupling coefficients based on above formalism, possibly pressure dependent neutron diffraction studies would prove useful in explaining this further.

**CONCLUSIONS**

In summary, using LSDA+U method as implemented in VASP, our density functional theory based calculations have explained the magnetic behavior of GFO vis-à-vis site disorder and subsequent magneto-structural coupling in multiferroic gallium ferrite. The calculated magnetic moments of Fe ions are in reasonable agreement with the experimental reports. We find that the cation site disorder not preferred in the ground state, Fe2-Ga2 site interchange is the most favored configuration in the disordered states. This appears to be driven by the thermal energy as the energy difference between two states is of the order of $k_BT$. An examination of the role of cation site disorder on magnetic structure of GFO showed modification of the local magnetic structure and altered magnetic moments of Fe ions at Ga site suggest that ferrimagnetism in GFO is solely due to site disorder. Further, an antiferromagentic coupling is predicted between Fe ions at Ga1 and Ga2 sites while Fe ions at Fe1 and Fe2 sites couple ferromagnetically. Finally, we find a compelling indication of, albeit somewhat weaker than compounds like $SrMnO_3$, magneto-structural coupling in GFO as changes in the spin arrangement led to significant variation in the structural distortion as well as magnetic moments.

## Acknowledgements

Authors thank Prof. M.K. Harbola and Prof. Rajeev Gupta (Physics Department, IIT Kanpur) for fruitful discussions and suggestions. Authors thank Ms. Somdutta Mukherjee for her help with the bulk magnetic measurements. AR thanks Ministry of Human Resources, Government of India for the financial support.

## References:

[1]     Prinz G A 1998 *Science* **282** 1660
[2]     Scott J F 2009 *Chem. Phys. Chem.* **10** 1761.
[3]     Scott J F 2007 *Nat. Mater.* **6** 256.
[4]     Baettig P and Spaldin N A 2005 *Appl. Phys. Lett.* **86** 012505.
[5]     Picozzi S, Yamauchi K, Sanyal B, Sergienko I A, Dagotto E. 2007 *Phys. Rev. Lett.* **99** 227201.
[6]     Bhattacharjee S, Bousquet E, Ghosez P 2009 *Phys. Rev. Lett.* **102** 117602.
[7]     Fennie C J 2008 Phys. Rev. Lett. **100** 167203.
[8]     Delaney K T, Mostovoy M, Spaldin N A 2009 Phys. Rev. Lett. **102** 157203.
[9]     Sosnowska I *et al*. 1982 *J. Phys. C: Solid State Phys.* **15** 4835.
[10]    Khomchenko V A, Kiselev D A, Bdikin I K, Shvartsman V V, Borisov P, Kleemann W, Vieira J M, Kholkin A L 2008 *Appl. Phys. Lett.* **93** 262905.
[11]    Das S R, Choudhary R N P, Bhattacharya P, Katiyar R S, Dutta P, Manivannan A, Seehra M S 2007 *J. Appl. Phys.* **101** 034104.
[12]    Kadomtseva A M, Popov Y F, Pyatakov A P, Vorob'ev G P, Zvezdin A K, Viehland D 2006 *Phase Transitions* **79** 1019.
[13]    Goswami S, Bhattacharya D, Choudhury P 2011 *J. Appl. Phys.* **109** 07D737.
[14]    Wang J, Neaton J B, Zheng H, Nagarajan V, Ogale S B, Liu B, Viehland D, Vaithyanathan V, Schlom D G, Waghmare U V, Spaldin N A, Rabe K M, Wuttig M, Ramesh R 2003 *Science* **299** 1719.
[15]    Wojdeł J C, Íñiguez J. 2009. *Phys. Rev. Lett.* **103** 267205.
[16]    Lee J H, Rabe K M 2010 *Phys. Rev. Lett.* **104** 207204.
[17]    Lueken H. A 2008 *Angew. Chem. Int. Edn.* **47** 8562.
[18]    Arima T *et al* 2004 *Phys. Rev. B* **70** 064426.

[19]     Remeika J P 1960 *J. Appl. Phys.* **31** S263.
[20]     Nowlin C H, Jones R V 1963 *J. Appl. Phys.* **34** 1262.
[21]     Kaneko Y, Arima T, He J P, Kumai R, Tokura Y 2004 *J. Magn.Magn.Mater.* **272-276** 555.
[22]     Mukherjee S, Gupta R, Garg A. 2011 *arXiv:1103.5541*.
[23]     Roy A, Mukherjee S, Gupta R, Auluck S, Prasad R, Garg A 2011 *J. Phys.: Condens. Matter* **23** 325902.
[24]     Han M J, Ozaki T, Yu J 2007 *Phys. Rev. B* **75** 060404.
[25]     Rado G T 1964 *Phys. Rev. Lett.* **13** 335.
[26]     Jones R O and Gunnarsson O 1989 *Rev. Mod. Phys.* **61** 689.
[27]     Kresse G and Furthmüller J 1996 *Phys. Rev. B* **54** 11169.
[28]     Blöchl P E 1994 *Phys. Rev. B* **50** 17953.
[29]     Anisimov V I, Aryasetiawan F, Lichtenstein A I 1997 *J. Phys.: Condens. Matter* **9** 767.
[30]     Kohn W and Sham L J 1965 *Phys. Rev.* **140** A1133.
[31]     Press W H, Flannery B P, Teukolsky S A, Vetterling W T. 1986 *Numerical Recipes: The Art of Scientific Computing* (New York: Cambridge University Press).
[32]     Monkhorst H J, Pack J D 1976 *Phys. Rev. B* **13** 5188.
[33]     Perdew J P *et al* 2008 *Phys. Rev. Lett.* **100** 136406
[34]     Goodenough J B 1955 *Phys. Rev.* **100** 564.
[35]     Goodenough J B 1958 *J. Phys. Chem. Solids* **6** 287.
[36]     Kim J Y, Koo T Y, Park J H 2006 *Phys. Rev. Lett.* **96** 047205.
[37]     Filippetti A, Hill N A 2000 *Phys. Rev. Lett.* **85** 5166.
[38]     Massidda S, Posternak M, Baldereschi A, Resta R 1999 *Phys. Rev. Lett.* **82** 430.
[39]     Corliss L M, Elliott N, Hastings J M 1960 *Phys. Rev.* **117** 929.
[40]     Singh D J, Pickett W E 1998 *Phys. Rev. B* **57** 88.

# List of Tables

**Table 1** Comparison of calculated and experimentally determined major bond angles in gallium ferrite unit-cell.

| Bond angle (all data are in °) | LSDA+U | GGA+U | Experiment |
|---|---|---|---|
| Fe1-O1-Fe2 | 168.54 | 168.33 | 166.11 |
| Fe1-O2-Fe2 | 103.06 | 103.30 | 99.33 |
| Fe1-O3-Fe2 | 123.13 | 123.07 | 121.00 |
| Fe1-O4-Fe2 | 143.06 | 143.90 | 143.22 |
| Fe1-O5-Fe2 | 126.23 | 126.21 | 126.20 |
| Fe1-O6-Fe2 | 174.46 | 174.63 | 173.25 |
| Fe1-O1-Ga2 | 166.08 | 165.81 | 164.08 |
| Ga1-O6-Fe2 | 124.12 | 124.06 | 122.57 |

**Table 2** Comparison of the calculated magnetization ($\mu_B$) data with previous calculations and experiments.

| Ion | LSDA+U without SOC[*] | LSDA+U with SOC[*] | (GGA+U) without SOC[*] | LSDA+U (without SOC)[†] | LSDA+U (with SOC)[†] | Experiment[††] |
|---|---|---|---|---|---|---|
| Ga1 | -0.01 | -0.01 | -0.01 | 0.01 | 0.01 | |
| Ga2 | 0.01 | 0.01 | 0.01 | 0.06 | 0.03 | |
| Fe1 | 4.05 | 4.05 | 4.12 | 5.02 | 4.27 | -3.9 |
| Fe2 | -4.04 | -4.04 | -4.12 | -5.11 | -4.34 | 4.5 |
| O1 | 0.06 | 0.06 | 0.06 | -0.03 to 0.06 | -0.07 to 0.06 | - |
| O2 | 0.00 | 0.00 | 0.00 | | | |
| O3 | 0.01 | 0.02 | 0.01 | | | |
| O4 | -0.05 | -0.05 | -0.05 | | | |
| O5 | 0.05 | 0.06 | 0.06 | | | |
| O6 | -0.07 | -0.07 | -0.07 | | | |

[*] **Present work;** [†] Han et al [21]; [††] Arima et al [1]

**Table 3** Calculated stress along crystallographic b-direction, maximum force on ions in the unit-cell and hydrostatic pressure on the unit-cell upon variation in spin configuration.

| | AFM-1 | AFM-2 | AFM-3 | AFM-4 |
|---|---|---|---|---|
| Stress ($N/m^2$) ×10$^{-4}$ | 42.7 | 45.2 | 45.2 | 65.9 |
| Force (max.) (eV/Å) | 0.0002 | 0.6400 | 0.6068 | 0.6500 |
| Hydrostatic Pressure (kB) | -0.27 | 3.96 | 10.16 | 9.37 |

# Figure Captions

Figure 1     Schematic crystal structure of GFO. The arrows indicate the direction of magnetic moment for Fe1 and Fe2 ions.

Figure 2     Change in the energy per unit-cell with respect to the ground state energy upon incorporating cation site disorder. The calculations were carried out using LSDA+U technique.

Figure 3     Changes in (a) the ground state lattice parameters and (b) magnetic moments for different spin configurations. Dashed lines are only as a guide to the eye.

Figure 4     Variation of magnetic moment as a function of applied hydrostatic stress on GFO unit-cell. Inset shows a plot of total energy for AFM-1 and AFM-2 structures versus applied pressure.

# **Figures**

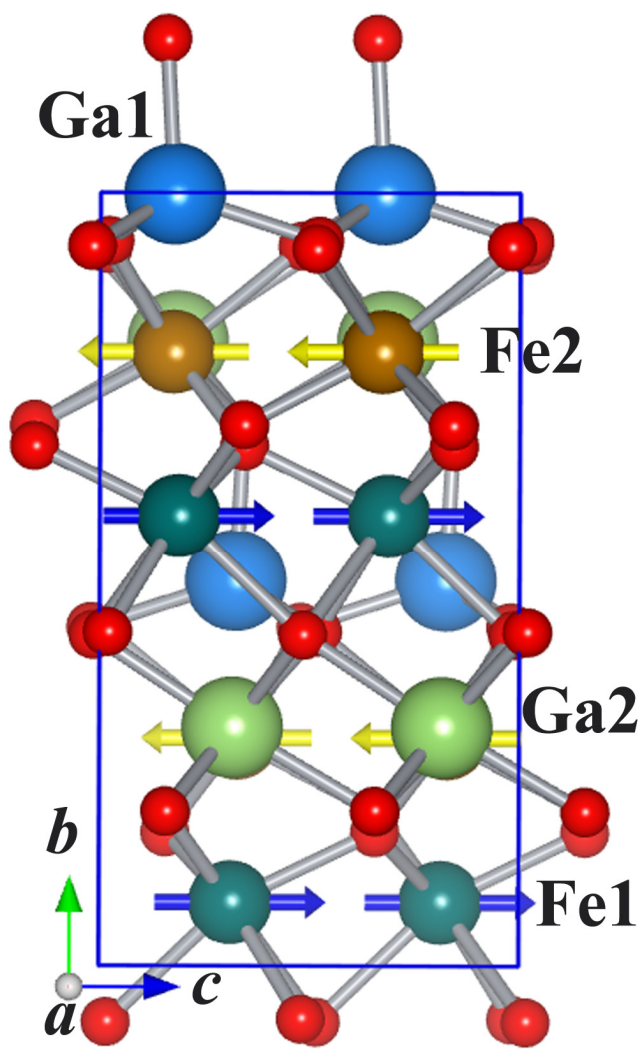

**Fig. 1, Roy** *et al*.

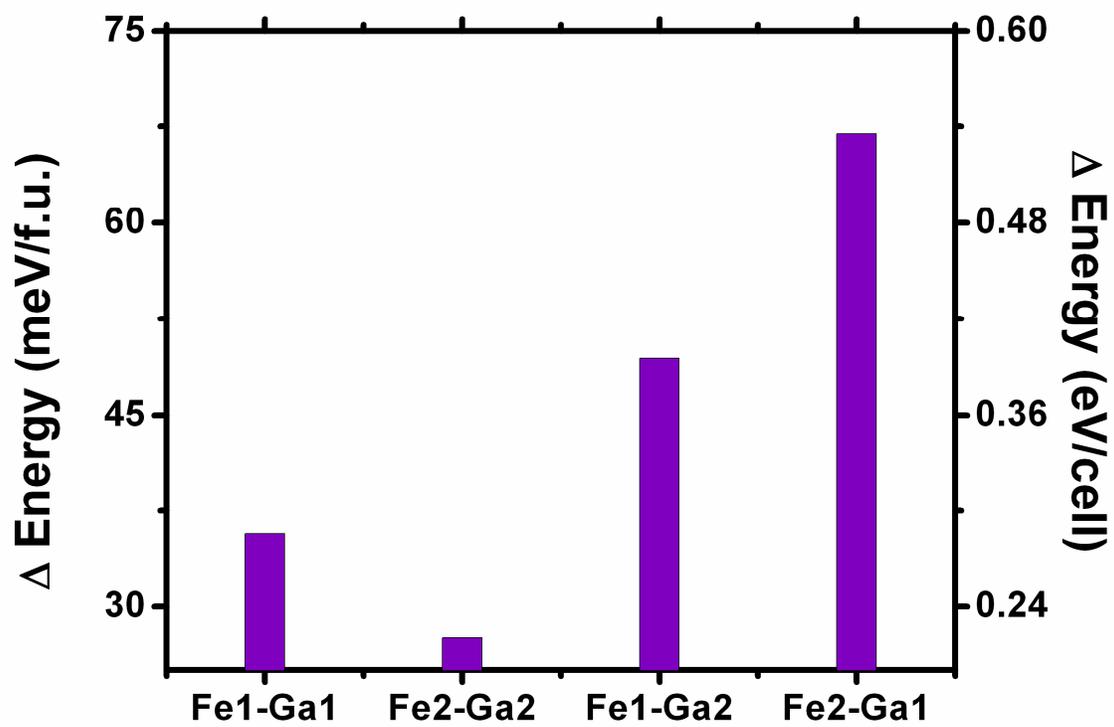

**Fig. 2, Roy** *et al*.

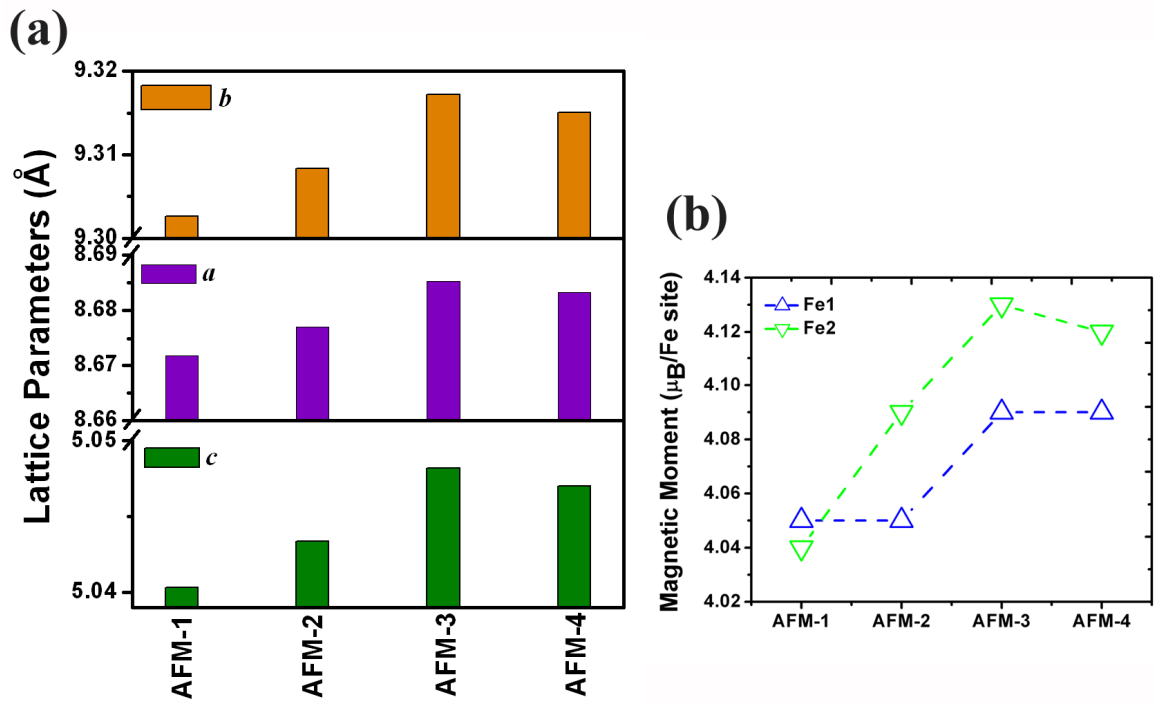

Fig. 3 Roy *et al*.

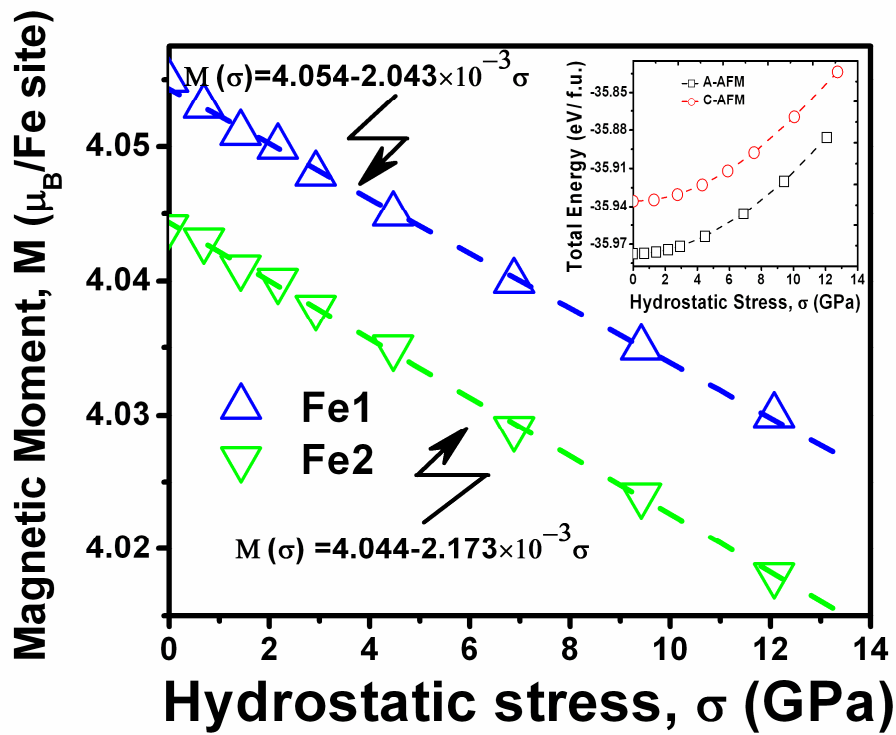

**Fig. 4** Roy *et al*.